\documentclass[%
 reprint,
 superscriptaddress,
 showkeys,
nofootinbib,
 amsmath,amssymb,
 aps,
 prd,
]{revtex4-2}

\usepackage{soul}
\usepackage{float}
\usepackage[caption=false]{subfig} 
\usepackage{gensymb}
\usepackage{graphicx}
\usepackage{dcolumn}
\usepackage{bm}
\usepackage[hypertexnames=true,colorlinks=true,urlcolor=blue,bookmarks=true,citecolor=blue,breaklinks=true]{hyperref}
\usepackage[dvipsnames]{xcolor}
\usepackage{booktabs}
\usepackage{makecell}
\usepackage{array}
\usepackage{multirow}
\newcolumntype{C}[1]{>{\centering\arraybackslash}p{#1}}

\begin{document}

\preprint{APS/123-QED}

\title{Toward More Realistic Machine-Learning Inference of the Dense-Matter Equation of State from Supernova Gravitational Waves}

\author{A. Akhmetali}
\affiliation{Energetic Cosmos Laboratory, Nazarbayev University, 010000 Astana, Kazakhstan} 
\affiliation{Department of Electronics and Astrophysics, Al-Farabi Kazakh National University, 050040 Almaty,
Kazakhstan} 

\author{Y. S. Abylkairov}
\email{sultan.abylkairov@nu.edu.kz} 
\affiliation{Energetic Cosmos Laboratory, Nazarbayev University, 010000 Astana, Kazakhstan} 
\affiliation{School of Artificial Intelligence and Data Science, Astana IT University, 020000 Astana, Kazakhstan}

\author{M. Zaidyn}
\affiliation{Department of Electronics and Astrophysics, Al-Farabi Kazakh National University, 050040 Almaty,
Kazakhstan} 

\author{A. Sakan}
\affiliation{Department of Electronics and Astrophysics, Al-Farabi Kazakh National University, 050040 Almaty,
Kazakhstan} 
\affiliation{Energetic Cosmos Laboratory, Nazarbayev University, 010000 Astana, Kazakhstan} 

\author{A. Zhunuskanov}
\affiliation{Department of Electronics and Astrophysics, Al-Farabi Kazakh National University, 050040 Almaty,
Kazakhstan} 

\author{N. Ussipov}
\email{ussipov.nurzhan@kaznu.kz}
\affiliation{Department of Electronics and Astrophysics, Al-Farabi Kazakh National University, 050040 Almaty,
Kazakhstan} 
\affiliation{Energetic Cosmos Laboratory, Nazarbayev University, 010000 Astana, Kazakhstan}

\author{J. A. Font}
\affiliation{Departamento de Astronomía y Astrofísica, Universitat de València, Avinguda Vicent Andrés Estellés 19, 46100 Burjassot (Valencia), Spain}
\affiliation{Observatori Astronòmic, Universitat de València, Catedrático José Beltrán 2, 46980, Paterna, Spain}

\author{A. Torres-Forné}
\affiliation{Departamento de Astronomía y Astrofísica, Universitat de València, Avinguda Vicent Andrés Estellés 19, 46100 Burjassot (Valencia), Spain}
\affiliation{Observatori Astronòmic, Universitat de València, Catedrático José Beltrán 2, 46980, Paterna, Spain}

\author{E. Abdikamalov}
\affiliation{Energetic Cosmos Laboratory, Nazarbayev University, 010000 Astana, Kazakhstan}
\affiliation{Department of Physics, Nazarbayev University, 010000 Astana, Kazakhstan}

\date{\today}

\begin{abstract}
Gravitational waves from core-collapse supernovae offer a unique probe of the equation of state (EOS) of dense nuclear matter. For rapidly rotating stars, previous machine-learning studies demonstrated promising EOS classification accuracy. However, these analyses relied on several simplifying assumptions. In this work, we relax three key assumptions. First, we include real detector noise. Second, we expand the analysis from a single progenitor model to four models spanning 12 to 40 solar masses, and for each mass we consider multiple rotational configurations, from slow to rapid. Third, we introduce uncertainty in the core bounce time of up to 20 ms, rather than assuming it is known precisely. We find that none of these effects significantly degrades EOS classification performance. Instead, the larger dataset associated with multiple progenitor models and noise realizations improves training and classification accuracy. This study is a step in a broader effort to progressively incorporate more realistic conditions into gravitational-wave inference for core-collapse supernovae.
\end{abstract}


\maketitle

\section{Introduction}
\label{sec:intro}

Gravitational waves (GWs) provide a powerful probe of astrophysical phenomena. In recent years, signals from compact binary mergers have been detected routinely \citep{Abac25GWTC4}, yielding important insights into the properties and physics of neutron stars and black holes. Another promising class of GW sources is core-collapse supernovae (CCSNe) \citep{abbott2020optically, szczepanczyk2023optically, powell22inferring}. For CCSNe occurring within our Galaxy, the resulting GW signals are expected to be detectable with current observatories \citep{gossan16observing, Szczepanczyk21Detecting}, while next-generation detectors may enable more detailed measurements and extend the reach to larger distances \citep{srivastava19detection}. Recent reviews of GW emission from CCSNe can be found in \citep{abdikamalov22gravitational, mezzacappa24gravitational, Mueller26GWREview}.

Massive stars reach the end of their lives when their cores exhaust nuclear fuel and can no longer support themselves against gravity. The core then collapses until the density becomes so high that repulsive nuclear forces halt the infall, launching a shock wave. This shock quickly loses energy and stalls, and must be revived in order to power a successful supernova explosion, leaving a neutron star as the remnant \citep{burrows93, Janka01Conditions, muller20hydrodynamics}. 

The newly formed proto-neutron star (PNS) is hot and cools by emitting neutrinos \citep{rampp00, oconnor13Progenitor}. Some of this energy is absorbed behind the stalled shock, heating the gas and stirring turbulence, which helps to energize the shock \citep{Buras06b, Bruenn16, Kotake18}. Additional support can come from large-scale instabilities such as the standing accretion shock instability (SASI) \citep{blondin03stability, foglizzo06neutrino, mueller:12} and from convective plumes originating in the nuclear-burning shells \citep{Couch15Three, Mueller17Supernova, Kazeroni20impact, Telman24Convective}.

In rapidly rotating progenitors, which likely represent only a small fraction of massive stars \citep{Heger05Presupernova, popov:12}, the explosion dynamics is considerably different. The PNS is born with substantial rotational kinetic energy that magnetic fields can tap to drive the explosion, producing magnetorotational outflows and collimated jets \citep{burrows:07b, moesta:14b, kuroda:20, obergaulinger:20, Bugli21Three}. Even jets that fail to break out of the stellar envelope can deposit enough energy to support the explosion \citep{Eisenberg22, Pais23choked}. At intermediate rotation rates, both neutrino heating and magnetic stresses may contribute to the outcome \citep{Takiwaki16Three, Summa18Rotation, abdikamalov2021, Buellet23Effect}.

While simulations continue to refine our theoretical models of core-collapse events \citep{muller20hydrodynamics}, the direct detection of neutrino and GW emissions from a nearby supernova would offer a transformative leap in data. Because GWs propagate directly from the engine of the star, they provide an unmediated probe into high-density regimes, potentially resolving long-standing questions regarding the explosion mechanisms of massive stars \cite{Powell24GW}. With advanced GW detectors approaching the sensitivity required to capture such signals \citep{gossan16observing, Szczepanczyk21Detecting}, developing reliable methods for identifying and studying CCSN GWs is increasingly important.

GWs arise from the multi-dimensional dynamics of the explosion engine, which depend strongly on rotation. In slowly rotating models, convection and the SASI excite PNS oscillations that dominate the signal \citep{Murphy09Model, Morozova18, Ott13General, Yakunin15GW, kuroda16, TorresForne19Universal, Mezzacappa23Core, Vartanyan23Gravitational, Sotani24Universality, Ehring26Gravitational, Lella26Gravitational}. In rapidly rotating cores, centrifugally-deformed bounce and subsequent ring-down of the PNS produces a GW burst \citep{ott12correlated, Fuller15SNseismology}. Differential rotation can drive long-lived, quasi-periodic, non-axisymmetric instabilities \citep{Scheidegger08, Shibagaki20new}. Additional low-frequency components may come from anisotropic neutrino emission \citep{mueller:97, Choi24GW}, asymmetric shock motion \citep{mueller12Parametrized, radice:19gw}, or jet activity \citep{Birnholtz13GW_jet, Gottlieb23Jetted, Soker23GWJJ}. In rotating models, GW emission may also be enhanced by the resonance between the fundamental quadrupolar oscillation mode of the PNS and the epicyclic oscillations at the boundary of the inner core~\citep{Cusinato:26}. 

Gravitational waves encode the internal evolution of their sources, allowing us to constrain the underlying supernova engine upon detection. Since rotation strongly influences core-collapse dynamics, GW observations can determine progenitor rotation \citep{abdikamalov:14, pajkos19, Akhmetali26PE} and discriminate between neutrino-driven and magnetorotational explosion models \citep{Logue12Inferring, Powell24Determining}. Beyond the explosion mechanism itself, GWs offer a way to inferring PNS parameters like mass and radius \citep{Bizouard21Inference, Sotani21Universal, Bruel23Inference, CasallasLagos23Characterizing}, and placing constrainghts on the nuclear equation of state (EOS) \citep{richers:17, edwards17, chao22determining, Wolfe23GW, Murphy24Dependence}.

In previous studies \citep{mitra24, Abylkairov24Evaluating, abylkairov2025assessing, Sakan25Probing}, we demonstrated the feasibility of using machine learning (ML) to infer the nuclear EOS from gravitational waves produced during rotating core bounce. While promising, those results relied on several simplifying assumptions: the use of simulated detector noise, a precisely known bounce time, and a dataset limited to the rotational configurations of a single progenitor model.

In this work, we relax these three assumptions to enhance the realism of our analysis. First, we incorporate real detector noise from the LIGO Hanford Observatory's latest O4a public release \citep{O4a}. Second, we expand our dataset to include four distinct progenitor models with zero-age main-sequence (ZAMS) masses ranging from $12$ to $40\,M_\odot$. While a progenitor can often be identified from pre-explosion images, its inner structure, which directly influences the GW signal, remains difficult to constrain \citep[e.g.,][]{Smartt09Progenitors, Sukhbold18High}. Thus, assessing the robustness of our results across a diverse range of progenitor models is essential. As in our prior work, we explore a wide spectrum of rotational configurations for each model. Third, we account for uncertainties in core-bounce timing. While the neutrino signal can estimate the bounce time to within $\sim10$ ms \citep{Pagliaroli09, Halzen09}, we allow for larger uncertainty of up to 20 ms. Electromagnetic observations are unlikely to improve this estimate because of the uncertain delay between core bounce and shock breakout.

The paper is structured as follows: Section~\ref{sec:methods} describes our dataset and methods. Section~\ref{sec:results} presents our results and evaluates the relative impact of noise, progenitor diversity, and bounce time uncertainty on EOS inference. Finally, Section~\ref{sec:conclusion} summarizes our main conclusions.

\section{Methods} 
\label{sec:methods}

We begin by describing the dataset, preprocessing steps, and the classification algorithm used to classify the different EOSs used in our study.

\subsection{Data}
\label{sec:data and noise}

\begin{figure}[t!]
\centering
\includegraphics[width=1\linewidth]{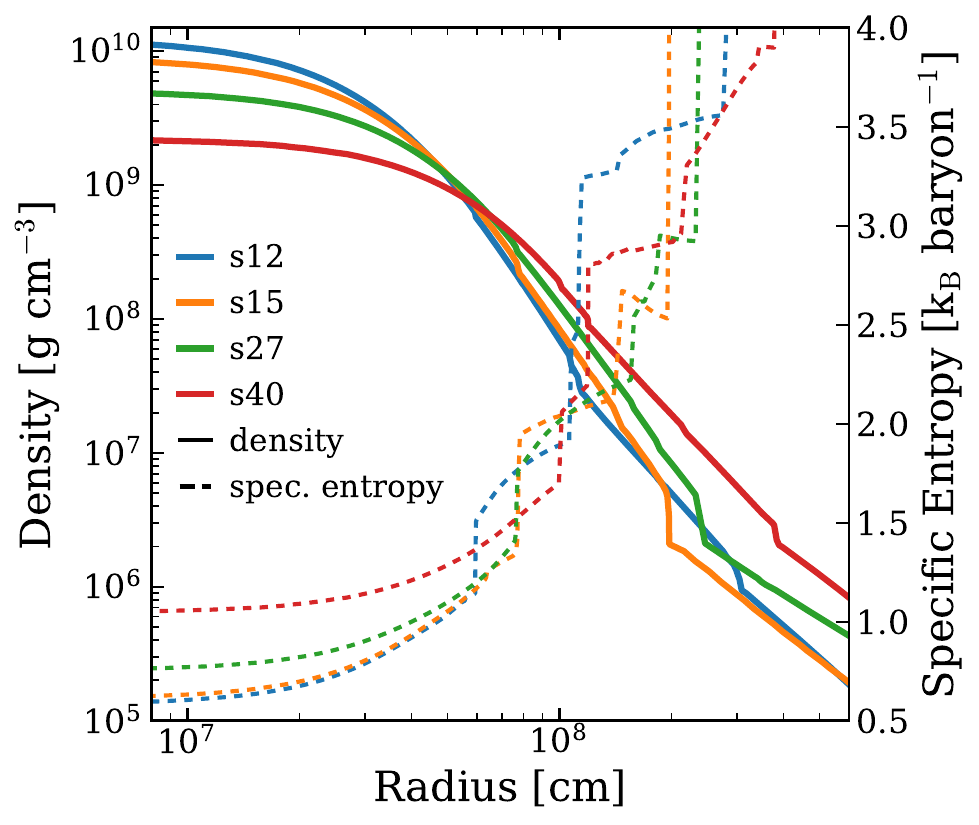}
    \caption{Radial profiles of density (solid lines) and specific entropy (dashed lines) for the 12 (blue), 15 (orange), 27 (green), and 40 (red) $M_\odot$ progenitor models prior to core collapse.}
     \label{fig:mass}
\end{figure}

The general relativistic {\tt CoCoNuT} code \citep{Dimmelmeier02a, dimmelmeier:05MdM} was used to obtain GW signals\footnote{The data are available at https://zenodo.org/records/17579189}. Since the details of the simulation setup (such as grid configuration or neutrino treatment) were presented in our previous works \citep{abdikamalov:14, richers:17, mitra23, mitra24}, we only highlight the key aspects here. Our simulations are axisymmetric, which is adequate during collapse, bounce, and the early post-bounce phase (see e.g.~\citep{ott:07cqg}). We employ the $Y_e(\rho)$ deleptonization scheme \citep{Liebendoerfer05Simple} during collapse and bounce, and switch to a leakage/heating scheme after bounce \citep{Ott13General}.

We consider four solar-metallicity progenitor models with ZAMS masses $12$ \cite{woosley:07}, $15$ \cite{heger:05}, $27$ \cite{whw:02}, and $40\,M_\odot$ \cite{woosley:07}. The higher-mass progenitors develop more massive, higher-entropy iron cores. At core collapse onset, the iron core masses are $1.3$, $1.4$, $1.5$, and $1.8\,M_\odot$, with central entropies of $0.60$, $0.62$, $0.77$, and $1.06\,k_\mathrm{B}$/baryon, respectively. Figure~\ref{fig:mass} shows the density and specific entropy as functions of radius. This set spans a diverse range of progenitor structures.

For each progenitor, we perform simulations with four EOSs: \texttt{SFHo} \cite{steiner:13b}, \texttt{LS220} \cite{lseos:91}, \texttt{HSDD2} \cite{hempel:10,hempel:12}, and \texttt{GShenFSU2.1} \cite{gshen:11b}. Because the focus of this work is on the effects of detector noise, progenitor variations, and bounce-time uncertainty, the particular choice of EOS does not influence how these factors affect the classification accuracy. The impact of EOS selection on classification accuracy has been examined in previous studies \citep{chao22determining, mitra24, abylkairov2025assessing}.

For each combination of mass and EOS, we simulated a number of rotational configurations, from slow to rapid. We quantify the impact of rotation by the ratio $T/|W|$, where $T$ denotes rotational kinetic energy and $W$ denotes gravitational binding energy. We consider $T/|W|$ values ranging from 0.02 to 0.21. This results in the number models per progenitor mass and per EOS ranging from 52 to 57. The number of rotation configurations varies because some EOS models do not undergo core collapse within the given simulation time limit. The total number of gravitational waveforms is 886, obtained by summing over all rotation configurations across each EOS and progenitor mass combination. All waveform data have a sampling rate of 4096 Hz, which matches Advanced LIGO's operational sampling rate.

We focus on the signal from $-2$ to $6$ ms relative to the time of bounce. The lower limit is motivated by the fact that there is little GW signal earlier than $\sim2$ ms before bounce. The upper limit is set because, after $\sim6$ ms post-bounce, the GW signal begins to include contributions from prompt convection, which is not accurately modeled in our axisymmetric simulations. While axisymmetry and the resolution we employ are adequate during bounce and the early post-bounce phase \citep{ott:07cqg}, convection is not well captured in 2D, particularly at the limited resolution used here \citep{Dolence13Dimensional, Mueller15dynamics, Raynaud20Magnetar}. Properly resolving the turbulent scales of prompt convection demands high-resolution 3D simulations, but the computational cost of generating the large dataset required for machine learning analysis remains prohibitive \citep{Radice16Neutrino}. Nevertheless, in Appendix~\ref{sec:prompt_convection}  we perform a test to assess how prompt convection (albeit modeled unphysically) might affect the EOS classification accuracy. We find that its impact is small, suggesting that the classifier primarily learns from the bounce and early post-bounce ring-down signal.

\subsection{Signal injection into detector noise}
\label{sec:injection}

For the background detector noise, we use publicly available LIGO Hanford data from the latest O4a public release~\citep{O4a}, starting at GPS time $t_{\rm GPS}=1378195220$ s, corresponding to early September 2023. The data are sampled at 4096~Hz, and a 1024~s segment of detector noise is used to inject the GW signals.

To generate a realistic training dataset, simulated gravitational wave signals are injected into the detector noise in the time domain. Prior to injection, each waveform is smoothly tapered using a Tukey window~\citep{tukey1967intoroduction} ($\alpha=0.1$) to suppress edge effects and reduce spectral leakage. 

\begin{figure}[t!]
\centering
\includegraphics[width=1\linewidth]{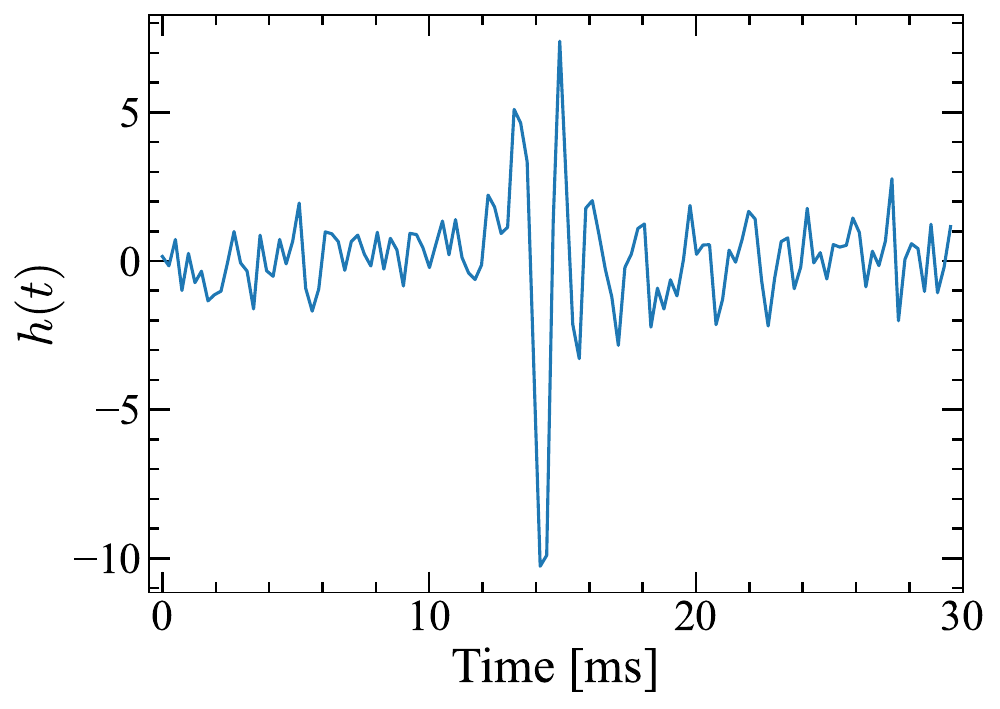}
    \caption{Example of a gravitational wave signal injected into O4a detector noise with $\mathrm{SNR}=20$.}
     \label{fig:Injected}
\end{figure}

The signal-to-noise ratio (SNR) of each raw waveform is computed using a global power spectral density (PSD) estimated from the full noise segment via Welch’s method~\citep{welch1967PSD}, according to \cite{flanagan98}
\begin{equation}
\text{SNR} = \sqrt{\int_0^\infty \frac{4|\tilde{h}(f)|^2}{S_n(f)} \, df} \,,
\end{equation}
where $\tilde{h}(f)$ is the Fourier transform of the signal and $S_n(f)$ is the one-sided PSD of the detector noise. This ensures consistent scaling of the injected signals. After injection, the combined signal is whitened, bandpass filtered between 20 and 2000 Hz, and notch filtered at the power line harmonics of 60, 120, and 240 Hz. An example of a injected signal is shown in Figure~\ref{fig:Injected}.

For the classification task, the resulting strain is segmented into 30 ms windows. As noted previously, the injected GW signals have a duration of approximately 8 ms. To account for the uncertainty in core bounce time, we inject the signal at a random point between the beginning of this 30 ms windows and time $\Delta t_\mathrm{b}$, which represents the uncertainty. We consider three value of bounce time uncertainty $\Delta t_\mathrm{b}$: 0, 10, and 20 ms.

\subsection{Classification algorithm}

To classify the EOS models using GW signals, we employ a support vector machine (SVM). This supervised ML algorithm demonstrated the best performance for similar tasks among the tested classical and deep learning models conducted by Abylkairov et al.~\citep{abylkairov2025assessing}. We use an rbf kernel SVM with a regularization parameter $C = 10$, selected through grid search cross validation to optimize performance on our dataset. We consider two input representations for the classifier: raw time series waveforms and their frequency domain counterparts.
For each selected SNR level, the model is trained on 80\% of the waveforms and evaluated on the remaining 20\%, with both training and testing samples consisting of signal injections into detector noise. 

To enforce a fixed target SNR across different rotational configurations, we rescale only the signal component while keeping the detector noise unchanged. This ensures that all injected waveforms attain the desired SNR under statistically consistent noise conditions. The signal-plus-noise waveform is constructed as
\begin{equation}
H = h \cdot \frac{\mathrm{SNR}{\mathrm{t}}}{\mathrm{SNR}{\mathrm{o}}} + n ,
\label{eq:2}
\end{equation}
where $\mathrm{SNR}{\mathrm{o}}$ is the original SNR of the injected signal in the given noise realization and $\mathrm{SNR}{\mathrm{t}}$ is the target SNR.
Eq.~\ref{eq:2} effectively corresponds to the optimally oriented case. For non-optimal orientations, the GW strain is additionally modulated by an inclination-dependent factor ($\propto \sin^2\theta$ for axisymmetric emission). However, this factor appears both in the waveform amplitude $h$ and in the measured $\mathrm{SNR}{\mathrm{o}}$, and is therefore canceled by the rescaling procedure. Consequently, all analyses presented in this work implicitly assume optimal source--detector orientations.

We evaluate the EOS classification performance using the accuracy metric, defined as
\begin{equation}
\text{Accuracy} = \frac{\text{Number of Correct Predictions}}{\text{Total Number of Predictions}} \,.
\end{equation}

To obtain statistically robust results, the evaluation procedure is repeated 50 times. In each iteration, a new random train--test split is generated. This repeated evaluation mitigates potential biases arising from a particular train--test partition. The final reported performance corresponds to the mean accuracy across all iterations, while the associated uncertainty is quantified by the standard deviation of these runs. Consequently, the reported results are effectively independent of any single random choice in the evaluation pipeline.

\section{Results} 
\label{sec:results}

We turn now to discuss how real detector noise, progenitor model variations, and bounce time uncertainty affect the EOS classification accuracy from core-collapse supernova gravitational waves.

\subsection{Impact of noise}
\label{sec:noise}

In our previous work, we analyzed the impact of simulated noise on the classification accuracy \citep{abylkairov2025assessing}. In the current work, we asses the impact of both real and simulated detector noise on EOS classification using our augmented dataset. The injection of GW signals into real O4a LIGO noise has been described in Section~\ref{sec:injection}. For simulated noise, we follow the same injection procedure, but the noise is generated as stationary, zero-mean Gaussian noise colored according to the detector PSD using the \texttt{PyCBC} package~\cite{pycbc}. In this test assessing the impact of noise, we assume that the bounce time is known precisely; the case of uncertain bounce time is considered below in Section~\ref{sec:bounce_time}. 

\begin{figure}[t!]
    \centering
    \includegraphics[width=1\linewidth]{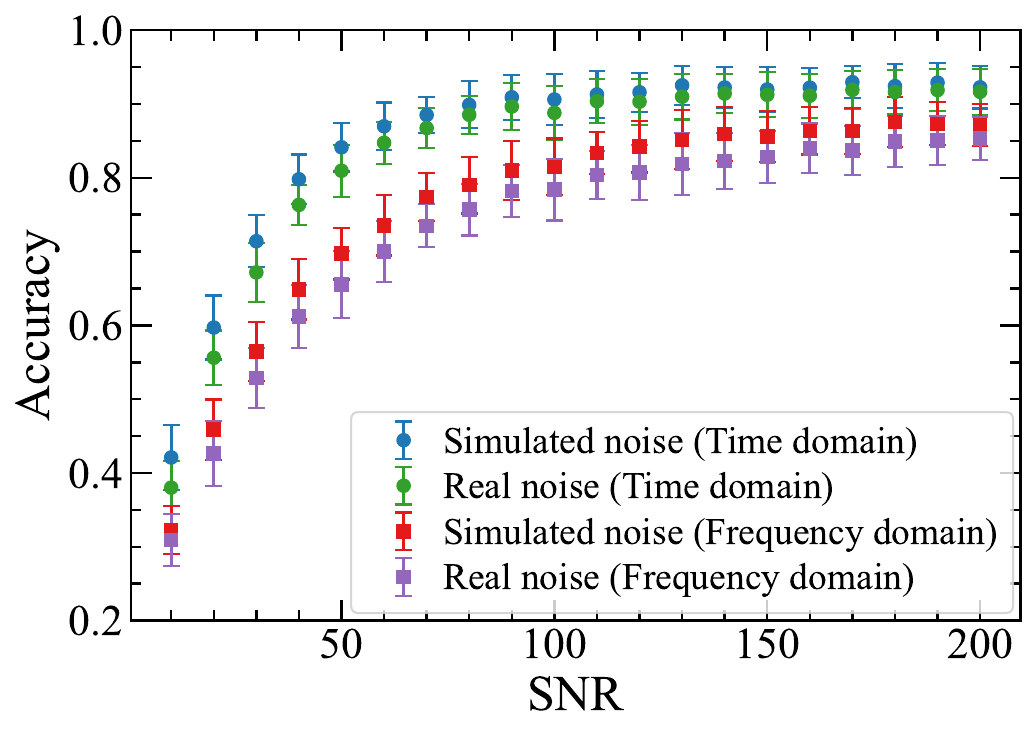}
    \caption{Accuracy as a function of SNR for real and simulated detector noises. Results are shown for two waveform representations in the time and frequency domains. At $\mathrm{SNR}=200$, the time series classifier achieves $92.3\%\pm2.9\%$ accuracy for simulated noise and $91.6\%\pm3.1\%$ for real noise, while the frequency domain classifier achieves $87.2\%\pm2.9\%$ for simulated noise and $85.3\%\pm2.9\%$ for real noise. The quoted uncertainties correspond to the $1\sigma$ standard deviation of the accuracies obtained from the 50 independent train--test splits described above.}
    \label{fig:noise}
\end{figure}

Figure~\ref{fig:noise} shows the classification accuracy as a function of signal-to-noise ratio (SNR) for signals injected into real and simulated noise. We analyze the data using two representations: the time-domain waveform and its frequency domain counterpart. In both cases, we find that injections into real and simulated detector noise yield similar classification accuracies. For example, at $\mathrm{SNR}=200$, for the time domain representation, the accuracies are $92.3\%\pm2.9\%$ and $87.2\%\pm2.9\%$ for real and simulated noise, respectively. For the frequency domain representation, the corresponding accuracies are $91.6\%\pm3.1\%$ and $85.3\%\pm2.9\%$, respectively. For reference, GW signals from Galactic CCSNe observed with second-generation detectors such as Advanced LIGO are expected to have signal-to-noise ratios of order $\mathrm{SNR}\sim10$–$30$, depending on the explosion dynamics and distance \citep{Szczepanczyk21Detecting, gossan16observing}. Third-generation detectors (e.g., Einstein Telescope, Cosmic Explorer) could reach $\mathrm{SNR}\gtrsim100$ for the same events \citep{srivastava19detection, abylkairov2025assessing}. The SNR range in Figure~\ref{fig:noise} thus reflect both current and next-generation detector sensitivities.

We find that the classification accuracy saturates at $\mathrm{SNR} \gtrsim 100$, with no significant gains at higher SNR. This behavior is likely driven by limitations in the size and diversity of the training dataset. Indeed, we observe that increasing the dataset size leads to improved classification performance (see next subsection). This trend indicates that further expanding the dataset (e.g., by more finely sampling rotational configurations in angular frequency) could yield additional gains in accuracy. However, the asymptotic performance limit remains unclear, as saturation with respect to dataset size has not yet been observed. 

These results indicate that the classifier performs similarly for signals injected into real and simulated detector noise. The small differences in accuracy likely arise from non-Gaussian and non-stationary features present in real detector noise that are not captured by the simulated Gaussian noise model. Nevertheless, the overall agreement suggests that simulated noise colored by the detector power spectral density provides a reasonable approximation for evaluating EOS classification performance in this and similar studies.

\subsection{Impact of progenitor diversity}
\label{sec:mass}

We now analyze how EOS classification depends on the diversity of progenitor models included in the dataset. Our dataset contains four progenitor models with masses $12\,M_\odot$, $15\,M_\odot$, $27\,M_\odot$, and $40\,M_\odot$ and we consider four different configurations. Configuration 1 includes a single progenitor mass, with results evaluated separately for each of the four progenitors and then combined using the root-mean-square (RMS) average. Configuration 2 includes all possible pairs of progenitor masses, configuration 3 includes all possible triplets, and configuration 4 includes all four progenitors simultaneously. For each configuration, we evaluate all possible combinations of progenitor masses of that size and report the classification accuracy as the mean and standard deviation over these combinations. Table~\ref{tab:unbalanced} summarizes the classification accuracy for these configurations.

\begin{table}[h]
\centering
\caption{Classification accuracy (\%) for various progenitor mass configurations at $\mathrm{SNR}=200$ and $\Delta t_\mathrm{b}=0$. Results are reported as the mean $\pm$ standard deviation over 50 iterations.} 
\label{tab:unbalanced}
\setlength{\tabcolsep}{3pt}
\begin{tabular}{lcccc}
\hline
\makecell{Configuration} & 1 & 2 & 3 & 4 \\
\hline
Frequency domain & $74.0\!\pm\!9.7$ & $80.4\!\pm\!5.9$ & $83.7\!\pm\!4.2$ & $85.3\!\pm\!2.9$ \\
Time domain  & $83.9\!\pm\!8.1$ & $87.7\!\pm\!4.9$ & $90.7\!\pm\!3.6$ & $91.6\!\pm\!3.1$ \\
\hline
\end{tabular}
\end{table}

\begin{table}[h]
\centering
\caption{Classification accuracy (\%) for various progenitor mass configurations of balanced dataset at $\mathrm{SNR}=200$ and $\Delta t_\mathrm{b}=0$. Results are reported as the mean $\pm$ standard deviation over 50 iterations.}
\label{tab:balanced}
\setlength{\tabcolsep}{3pt}
\begin{tabular}{lcccc}
\hline
\makecell{Configuration} & 1 & 2 & 3 & 4 \\
\hline
Frequency domain & $73.1\!\pm\!8.6$ & $68.6\!\pm\!9.3$ & $69.3\!\pm\!7.5$ & $66.8\!\pm\!12.2$ \\
Time domain  & $83.9\!\pm\!8.1$ & $80.3\!\pm\!7.9$ & $78.0\!\pm\!8.9$ & $79.7\!\pm\!9.0$ \\
\hline
\end{tabular}
\end{table}

As we can see, classification accuracy increases—and standard deviation decreases—as additional progenitor masses are included. For the frequency domain representation, the accuracy rises from $74.0\%\pm9.7\%$ for a single progenitor to $85.3\%\pm2.9\%$ when all four progenitors are used. Similarly, for time domain representation, the accuracy increases from $83.9\%\pm8.1$ to $91.6\%\pm3.1\%$. While adding progenitor diversity should complicate the classification task, this upward trend stems from two primary factors. First, progenitors of different masses yield similar bounce GW signals if they possess the same specific angular momentum for a given enclosed mass. \citep{mueller:09phd, ott12correlated, mitra23}. Second, the inclusion of more progenitor models expands the training dataset. Because machine learning models typically generalize better when trained on larger datasets, the overall classification performance increases.

\begin{figure*}[ht]
\centering
\includegraphics[width=0.95\linewidth]{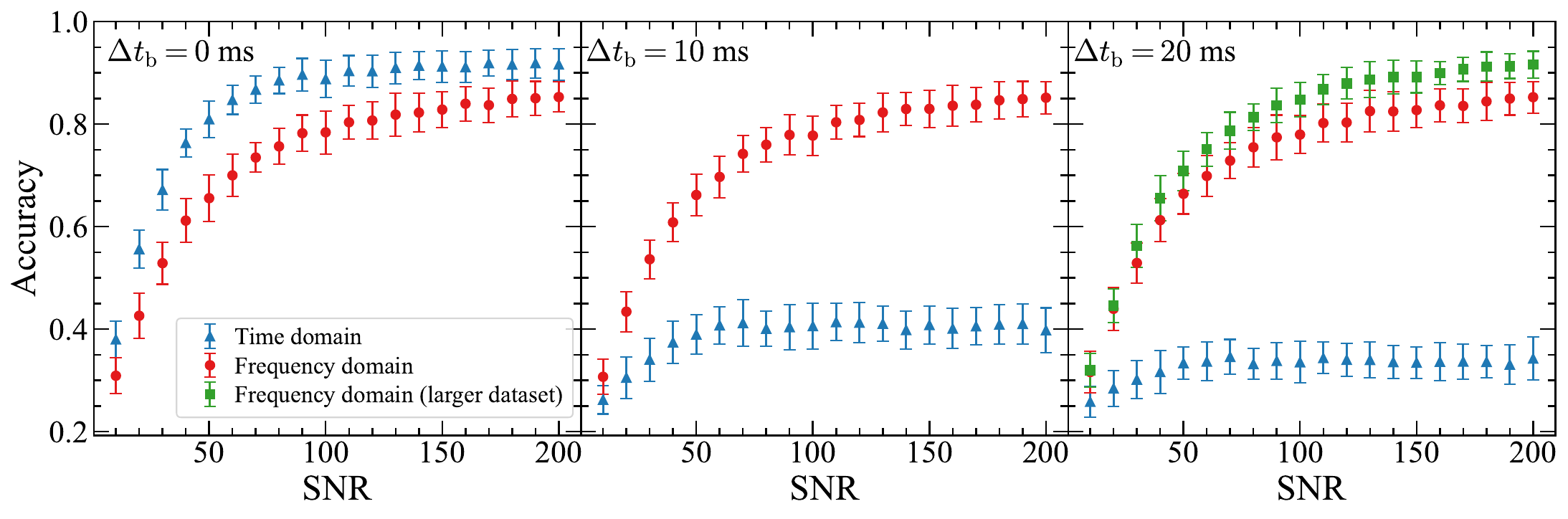}
    \caption{EOS classification accuracy as a function of SNR.
    Blue triangles represent classification results in the time domain, red circles indicate classification in the frequency domain, and green rectangles show frequency-domain classification using an expanded dataset. The left, middle, and right panels correspond to the case with 0, 10, and 20 ms bounce time uncertainty cases. The title of each sub-panel indicates the assumed bounce time uncertainty $\Delta t_\mathrm{b}$.}
     \label{fig:accuracies}
\end{figure*}

To test this hypothesis, we repeat the same calculations using a balanced dataset. To construct this dataset, we randomly subsample signals in each class so that the total number of training examples is equal across all scenarios, regardless of the number of progenitors. Specifically, the balanced dataset contains 220 signals for each configuration. This ensures that differences in classification accuracy are not driven by unequal dataset sizes.

Table~\ref{tab:balanced} presents the results for the balanced dataset. As expected, in this case, the trend is opposite: classification accuracy decreases as additional progenitor models are included, and the standard deviation also increases. For the frequency domain representation, accuracy drops from $73.1\%\pm8.6\%$ for a single progenitor to $66.8\%\pm12.2\%$ when all four progenitors are included. For the time domain representation, accuracy decreases from $83.9\%\pm8.1\%$ to $79.7\%\pm9.0\%$ across the same range. 

Time domain representation consistently achieve accuracies a few percentage points higher than those based on frequency domain across all configurations (cf.~Tables \ref{tab:unbalanced}-\ref{tab:balanced}). This likely reflects their ability to preserve the sharp, short-duration structures of the bounce signal, which are partially smoothed out in the Fourier amplitude spectrum. As we will see below, however, the situation changes drastically once bounce time uncertainty is introduced.

Overall, the comparison highlights that dataset size plays a more important role in classification performance than progenitor diversity. In the unbalanced dataset, the observed improvement in accuracy and reduction in standard deviation with additional progenitors is primarily driven by the larger amount of training data, which helps the classifier better learn EOS specific features. When the dataset is balanced and the number of training examples is kept fixed, classification accuracy decreases only modestly as additional progenitor masses are included. This indicates that, although different progenitors introduce some variability in the gravitational wave signals, these differences do not significantly degrade the model’s ability to identify EOS dependent features.  

\subsection{Impact of bounce time uncertainty}
\label{sec:bounce_time}

We now analyze how bounce time uncertainty affects EOS classification. Figure~\ref{fig:accuracies} shows the classification accuracy across different SNR levels for time and frequency domain representations. The left, middle, and right panels correspond to bounce time uncertainty of 0, 10, and 20 ms, respectively. As mentioned before, when bounce time is known precisely, both time and frequency domain representations achieve high accuracies of $91.6\% \pm 3.1\%$ and $85.3\% \pm 2.9\%$ at $\mathrm{SNR}=200$, respectively.

The results change markedly when bounce time uncertainty is introduced. As shown in the middle and right panels of Figure~\ref{fig:accuracies}, the time domain classifier struggles, achieving only $39.8\%$ at $\Delta t_{b}=10$ ms and $34.3\%$ at $\Delta t_{b}=20$ ms. This behavior is expected, as random shifts in the bounce time obscure the bounce and post-bounce features that are essential for distinguishing among EOSs \cite{mitra23, Abylkairov24Evaluating}. 

In contrast, the frequency-domain classifier remains largely unaffected by bounce time uncertainty. Even for $\Delta t_\mathrm{b}=20$ ms, it maintains an accuracy of $85.3\% \pm 3.1\%$ at $\mathrm{SNR}=200$, indicating that the spectral features used for classification are robust against time shifts in the signal.

Moreover, we explore whether increasing the size of the training dataset can further improve the performance of the frequency-domain classifier. To this end, we repeat the training while injecting the signals with a time step of 2\,ms when applying the bounce time shifts. This procedure effectively increases the number of training samples elevenfold. As shown in the right panel of Figure~\ref{fig:accuracies} (green rectangles), the classification accuracy improves even further in this case, reaching $91.6\% \pm 2.7\%$ at $\mathrm{SNR}=200$. This result suggests that the frequency-domain classifier can benefit from larger training sets while remaining robust to bounce time uncertainty.


These results show that bounce-time uncertainty affects the two approaches differently. Time-series classification is sensitive to shifts in signal timing: when the temporal features become misaligned between the training and testing samples, the classification accuracy decreases significantly. In contrast, the frequency-domain representation depends primarily on the spectral content of the signal rather than its exact temporal position. As a result, the frequency domain classifier maintains high performance even for large timing uncertainties.

There is, however, one caveat. In our analysis we use GW signals around bounce time and exclude contribution of the prompt convection. In Appendix~\ref{sec:prompt_convection}, we perform additional tests by including the contribution of the prompt convection (albeit in an approximate way) and find that it does not affect significantly the classification accuracy, suggesting that the classifier mainly learns EOS features from the bounce and early post-bounce ring-down signal. See Appendix~\ref{sec:prompt_convection} for further details. 

\section{Conclusion} 
\label{sec:conclusion}

In this work, we have investigated the prospect of classifying the nuclear equation of state using gravitational-wave signals under more realistic observational conditions than those employed in \citep{mitra24, Abylkairov24Evaluating, abylkairov2025assessing}. Compared to our previous studies, we improve the realism in three ways: by using O4a LIGO detector noise, incorporating more progenitor models, and accounting for uncertainty in the bounce time.

First, we find that the inclusion of real detector noise degrades the classification accuracy by only a few percent (cf. Section~\ref{sec:noise}). Second, our results show that variations in the progenitor mass do not significantly affect the EOS classification performance. On the contrary, because this introduces additional models into the dataset, increasing the dataset size through progenitor mass variation improves the overall training and classification accuracy (cf. Section~\ref{sec:mass}). Finally, we evaluate the machine-learning performance under bounce time uncertainties of up to 20 ms. We perform the analysis in both the time and frequency domains. The frequency-domain representation consistently achieves high classification accuracy across all bounce time uncertainty cases. In contrast, the time-domain representation maintains high accuracy only when the bounce time is known precisely, which is difficult to achieve in realistic scenarios (cf. Section~\ref{sec:bounce_time}). 

These results indicate that real detector noise, progenitor variations, and bounce-time uncertainty do not significantly affect the EOS classification accuracy. We also observe a clear trend that the accuracy improves as the size of the training dataset increases. While these results are encouraging, several limitations remain that prevent more definitive conclusions. 
In particular, our analysis is based on a discrete set of representative EOS models, and the machine-learning task is therefore formulated as a classification problem among these EOSs. While the selected EOSs span a range of neutron-star properties, they do not fully capture the uncertainty in the high-density matter EOS. A more general approach would be to construct a parameterized family of EOS models spanning the relevant high-density nuclear matter properties, and to treat the inference problem as a regression task aimed at recovering the underlying EOS parameters directly from the signal.
Moreover, our simulations do not fully capture physical processes that occur at later stages of the evolution, such as convection and anisotropic neutrino emission. Finally, we assume optimal source and detector orientations. A more realistic treatment should also include detector antenna responses, polarization mixing, and off-axis source configurations, requiring a detailed analysis with networks of current and future GW detectors. We plan to address these limitations systematically in future work.

\appendix

\section{Contribution of prompt convection}
\label{sec:prompt_convection}

In analysis presented in the main body of this article, we focused on short GW signals of 8 ms duration around core bounce. Those are dominated by the bounce and early ring-down phases of the PNS. At later times, however, the GW signal contains additional physical contributions, such as prompt convection. To assess whether the presence of this component affects EOS classification, we perform an additional experiment using GW signals from the CCSN simulations of \citet{richers:17}, which follow the evolution up to 50 ms after bounce and thus include the contribution of the prompt convection.

\begin{figure}[t]
    \centering
    \includegraphics[width=1\linewidth]{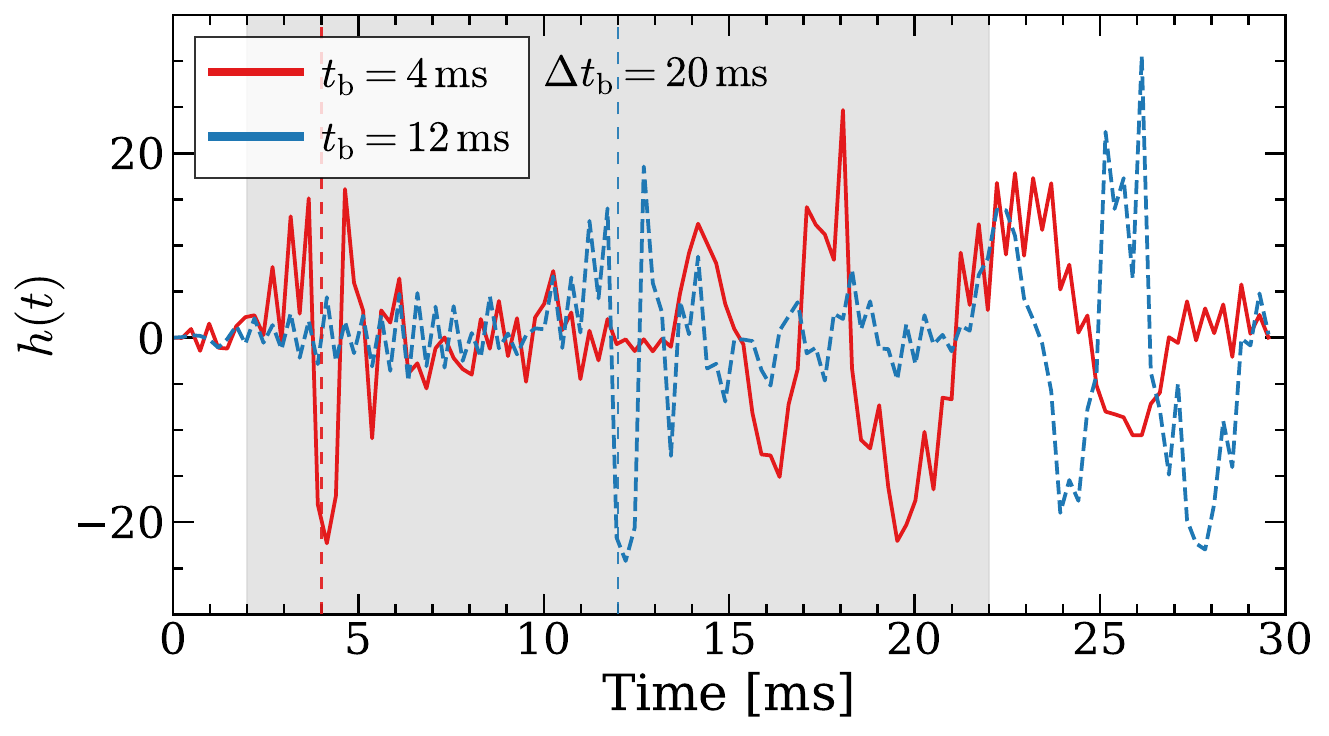}
    \caption{Examples of two GW signals with two different bounce time assignments. The bounce time uncertainty is $\Delta t_\mathrm{b}=20$ ms. Red and blue lines show the same signal with different assigned bounce times, illustrating that variations arise from both shifted bounce times and intrinsic waveform differences. The gray shaded region indicates the bounce time possible position interval.}
    \label{fig:bounce_uncertainty}
\end{figure}

The original  dataset from \citep{richers:17} contains 18 EOS models. To maintain a consistent classification task, we select the four EOSs that are also present in our dataset. We inject each signal to random 30 ms noise window by randomly selecting the bounce time from 2 ms to $\Delta t_\mathrm{b} + 2 $ ms, where $\Delta t_\mathrm{b}$ is the bounce time uncertainty. See Figure~\ref{fig:bounce_uncertainty} for examples with two bounce times at 4 and 12 ms for $\Delta t_\mathrm{b} = 20 $ ms.

\begin{figure}[t]
\centering
\includegraphics[width=1\linewidth]{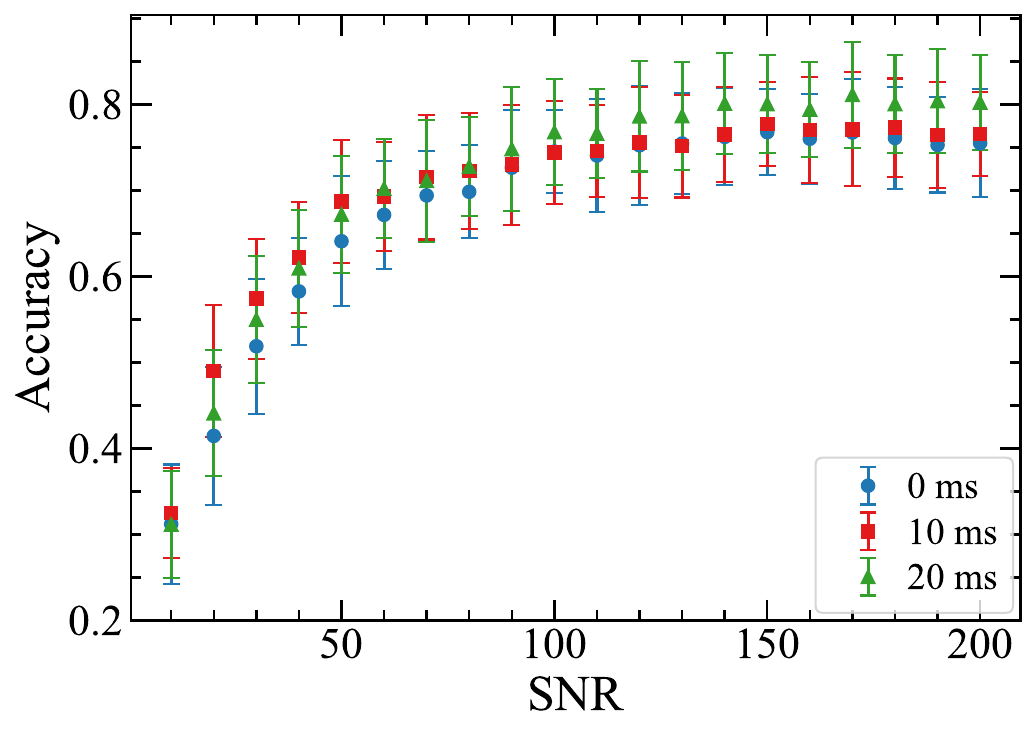}
    \caption{EOS classification accuracy as a function of SNR for GW data that contains the contribution of prompt convection. Legends corresponds to different values of bounce time uncertainty $\Delta t_\mathrm{b}$.}
     \label{fig:prompt_convection}
\end{figure}

The classification results for this dataset are shown in Figure~\ref{fig:prompt_convection}. The classifier achieves $80.2\% \pm 5.5$ classification accuracy for $\Delta t_\mathrm{b} = 20$ ms at $\mathrm{SNR}=200$. This is $11.4\%$ lower than the corresponding accuracy for our dataset without prompt convection ($91.6 \pm 2.6\%$). This reduction is expected, since the analysis window may now include different portions of the post-bounce evolution, where prompt convection contributes stochastic variations to the waveform. Such variability increases the diversity of signal realizations and makes EOS classification more challenging.

\begin{figure}[t]
\centering
\includegraphics[width=1\linewidth]{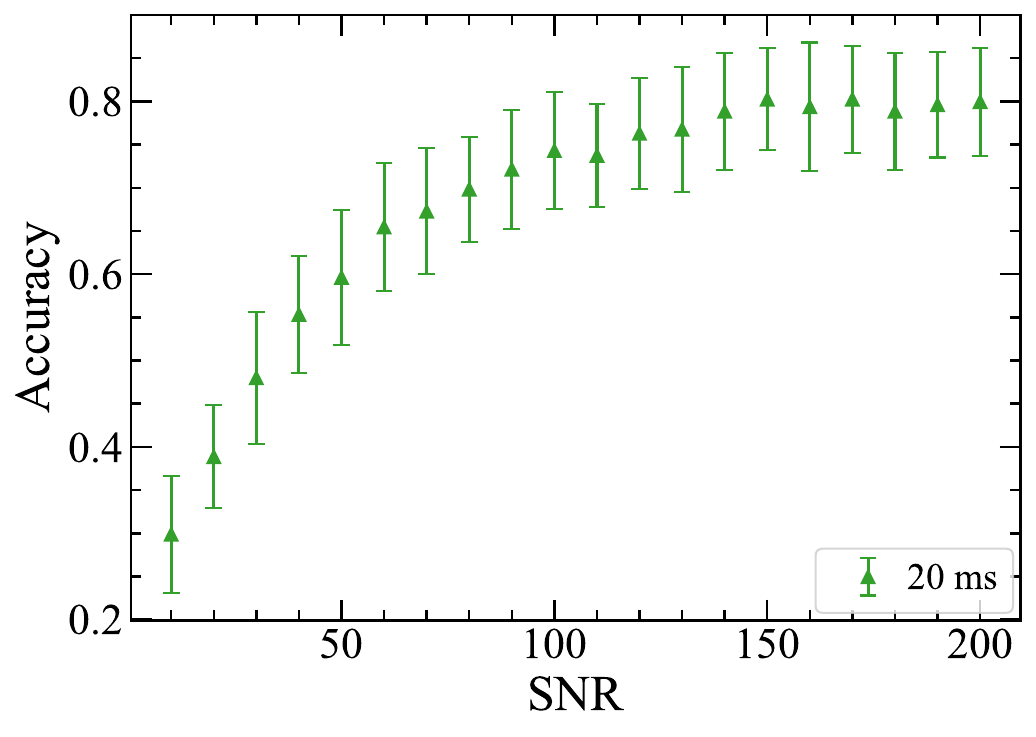}
    \caption{Overfitting test with $\Delta t_\mathrm{b}=20$ ms. In this test, we restrict $\Delta t_\mathrm{b} = [18,20]$ ms in our test data to exclude parts of the signals dominated by prompt convection. This test verifies that our classifies learns from the bounce and early post-bounce PNS oscillations as opposed to the convection signal.}
     \label{fig:prompt_convection_overfitting}
\end{figure}

Because convection is inherently stochastic, each simulation produces only a single realization of this process. As a result, there is a risk that the classifier could learn features specific to that particular realization of the convective signal rather than properties robustly related to the EOS. This raises an important question: can we be sure that the classifier is not relying on the stochastic component instead of the bounce signal itself?
To address this, we perform an additional test in which the training procedure remains unchanged, but for testing we use data with the bounce time selected in the interval $[18,20]$ ms. This constraint ensures that most of the prompt convection signal lies outside our 30 ms window and is therefore not included in the classification test.

Figure~\ref{fig:prompt_convection_overfitting} shows that the classification performance remains essentially unchanged when the test data are restricted to bounce times between $18$ and $20$ ms. The accuracy for this restricted dataset is $79.6\% \pm 6.2\%$, compared to $80.2\% \pm 5.5\%$ for the original dataset that includes prompt convection. This demonstrates that excluding the portion of the signal potentially affected by stochastic variations from prompt convection has a negligible effect on the classifier’s predictions. Thus we conclude that the model primarily relies on physically meaningful features associated with the bounce and early post-bounce PNS oscillations, which encode EOS-dependent information.

\begin{acknowledgments}
This research was funded by the Science Committee of the Ministry of Science and Higher Education of the Republic of Kazakhstan (Grant No. AP26103591). EA is partially supported by the Nazarbayev University Faculty Development Competitive Research Grant Program (no. 040225FD4713).
JAF and ATF are supported by the Spanish Agencia Estatal de Investigación (grant PID2024-159689NB-C21) funded by 
MICIU/AEI/10.13039/501100011033 and by FEDER / EU, and by the Generalitat Valenciana (Prometeo Excellence Programme grant CIPROM/2022/49).

AI-based language tools were used to improve the clarity and grammar of the manuscript.

This research has made use of data or software obtained from the Gravitational Wave Open Science Center (gwosc.org), a service of the LIGO Scientific Collaboration, the Virgo Collaboration, and KAGRA.
\end{acknowledgments}

\bibliography{gw_sn}

\end{document}